\documentclass[a4paper,11pt]{article}

\usepackage{jcappub} 

\title{Two-fluid model of the pulsar magnetosphere represented as an axisymmetric force-free dipole}

\author{S. A. Petrova}

\affiliation{Institute of Radio Astronomy of the NAS of Ukraine,\\Mystetstv Str., 4, Kharkiv 61002, Ukraine}

\emailAdd{petrova@rian.kharkov.ua}

\abstract{Based on the exact dipolar solution of the pulsar equation the self-consistent two-fluid model of the pulsar magnetosphere is developed. We concentrate on the low-mass limit of the model, taking into account the radiation damping. As a result, we obtain the particle distributions sustaining the dipolar force-free configuration of the pulsar magnetosphere in case of a slight velocity shear of the electron and positron components. Over most part of the force-free region, the particles follow the poloidal magnetic field lines, with the azimuthal velocities being small. Close to the Y-point, however, the particle motion is chiefly azimuthal and the Lorentz-factor grows unrestrictedly. This may result in the very-high-energy emission from the vicinity of the Y-point and may also imply the magnetocentrifugal formation of a jet. As for the first-order quantities, the longitudinal accelerating electric field is found to change the sign, hinting at coexistence of the polar and outer gaps. Besides that, the components of the plasma conductivity tensor are derived and the low-mass analogue of the pulsar equation is formulated as well.}


\begin{document}
\maketitle
\flushbottom

\section{Introduction}
Pulsars are known to be rotating magnetized neutron stars, with the stellar magnetic field being roughly dipolar (for a more advanced conception of the magnetic field structure of the neutron star see, e.g., \cite{pons09,b15}). The pulsar magnetosphere, however, cannot be adequately represented by the vacuum rotating dipole due to the presence of a dense electron-positron plasma which should affect the electromagnetic fields. The self-consistent treatment of the currents and fields in the pulsar magnetosphere proceeds from the model of an ideal axisymmetric force-free rotator containing enough plasma to screen the accelerating longitudinal electric field and to provide the electromagnetic force balance. In this case, the pulsar magnetosphere is described by the well-known pulsar equation \cite{m73,sw73}. In its most general form \cite{o74}, the pulsar equation relates the magnetic flux, the poloidal current, and the transverse potential drop causing differential rotation of the magnetosphere.

The first exact solution of the pulsar equation was guessed at once \cite{m73}, and for long years it was thought to be indirectly relevant to the pulsar magnetosphere. The solution corresponds to the monopolar magnetic field rigidly corotating with the neutron star. Although the monopolar solution allows straightforward generalization to the arbitrary profile of the differential rotation velocity (e.g., \cite{p13}), this was used in the literature only occasionally (e.g., \cite{c05}), mainly because of the lack of an idea as to the realistic form of such a profile. Of course, direct application of the monopolar solution to the neutron star would be somewhat unnatural, but its extensions to the cases of a split \cite{m91} and offset \cite{p13} monopole are supposed to mimic the magnetospheric structure far from the neutron star. It is the split-monopole picture that prompted the set of boundary conditions used traditionally in the numerical simulations of the pulsar magnetosphere \cite{ckf99}: the magnetic field is assumed to be dipolar at the neutron star, monopolar-like at infinity and continuous across the light cylinder. The subsequent numerical studies proved the uniqueness of such a solution and made a number of advances (see \cite{petri16} for a recent review).

Being merely hinted by the exact split-monopole solution and based on the general considerations as to smoothness of the plasma flow, the customary set of boundary conditions \cite{ckf99} involved in the numerical studies of the pulsar magnetosphere for long time escaped physical corroboration. As was shown in \cite{p12}, the quasi-monopolar structure at infinity, smooth passage through the light cylinder and zero poloidal current at the magnetic axis exclude the purely dipolar condition at the stellar surface. Recently, the ab initio simulations of the pair cascade in the pulsar force-free magnetosphere have demonstrated that the customary set of boundary conditions does not provide intense enough plasma production to supply the force-free regime over most part of the magnetosphere \cite{abinit1,abinit2}. This questioned the existing magnetospheric picture as a whole. Scarce alternative magnetospheric pictures suggesting modifications in the equatorial \cite{gruz11,p13} and axial \cite{sl90,l06,t14} regions are conjectural as well.

Obviously, the pulsar equation should have other exact solutions which may also give valuable insights as to the boundary conditions for the pulsar magnetosphere. Recently we have serendipitously found the exact dipolar solution of the pulsar equation and constructed the magnetospheric model on its basis \cite{p16}. It appears to differ substantially from what is commonly thought of the pulsar force-free magnetosphere (see also Sect.~2 below for details) and, surprisingly, seems to have some observational support. In the present paper, we elaborate our model of an axisymmetric force-free dipole by finding the particle distributions which may sustain this magnetospheric configuration.

The force-free treatment and the resultant pulsar equation involve only the plasma charge and current densities, leaving aside the particle distribution functions. For the classical monopolar case, however, the degenerate picture is well known \cite{m73}: the particles stream along the poloidal magnetic field lines at a speed of light. Later on this was generalized for arbitrary force-free electromagnetic fields \cite{gruz08}, in which case the particle speed-of-light motion is directed so as to preclude the radiation damping.

Of course, for the sake of studying the physical processes  and the resultant emission in the pulsar magnetosphere, the realistic non-degenerate particle distributions are of much more interest. They can be obtained within the framework of the self-consistent two-fluid model. Although such a model is generally too complicated and allows only simplistic numerical illustrations \cite{ko09}, the low-mass limit of the model (which is indeed suitable for pulsars) may well be addressed analytically. For the monopolar case, this was done in \cite{b00,p15}. In the present paper, we generalize our previous formalism \cite{p15} for an arbitrary force-free configuration and apply it to the axisymmetric force-free dipole in order to find the particle distribution functions and trajectories as well as the components of the plasma conductivity tensor in the realistic pulsar magnetosphere.

Note that the classical quasi-monopolar numerically constructed models proceed from the minimal set of boundary conditions providing smoothness of the solution over the whole space. Within the framework of such a problem, the global force-free solution obtained is known to exclude any physical processes in the magnetosphere. In our case, it is the purely dipolar analytic solution that dictates the boundary conditions, the boundaries themselves and, ultimately, the corresponding magnetospheric physics.
As is shown in \cite{p16} (see also Sect.~2 and Fig.~\ref{f1} below), in the purely dipolar case, the force-free regime holds only in a certain region inside the magnetosphere rather than over the whole space. But it is the force-free region that is believed to be responsible for the pulsar radio emission and most probably for the non-thermal high-energy emission as well for geometrical and also for physical reasons, since the energetics of these emissions is much lower than that of the particle flow. Furthermore, specification of the physical conditions at the boundary of the force-free zone should provide a proper basis for studying the physics of the neighbouring regions of the magnetosphere and the processes at the boundaries themselves. Thus, our present research has strong enough motivations.

The plan of the paper is as follows. In Sect.~2, we sketch a general scheme of the pulsar magnetosphere in the form of an axisymmetric force-free dipole. Sect.~3 is devoted to the two-fluid model with the particle inertia being ignored. The low-mass limit of the two-fluid model is addressed in Sect.~4. Our results are discussed in Sect.~5 and briefly summarized in Sect.~6.

\section{General picture of the dipolar force-free magnetosphere}
Let us consider the pulsar magnetosphere based on the model of an axisymmetric force-free dipole. As is found in \cite{p16}, the magnetic flux function $f$ in the form of a pure dipole,
\begin{equation}
f=\sin^2\theta/r\,,
\label{eq1}
\end{equation}
(where $r,\theta$ are the coordinates in the spherical system with the axis along the pulsar axis) is the exact solution of the pulsar equation, with the corresponding poloidal current function $g(f)$ and angular rotation velocity $\mu(f)$ being
\begin{equation}
g=\sqrt{\frac{4C_1^2}{f^2}+C_2}\,,\quad \mu=\frac{C_1}{f^2}\,,
\label{eq2}
\end{equation}
where $C_1$ and $C_2$ are arbitrary constants. Note that throughout the paper we take $c=e=\Omega_*=M=1$, where $c$ is the speed of light, $e$ the electron charge, $\Omega_*$ and $M$ the rotational velocity and magnetic moment of the neutron star. Of course, in the closest vicinity of the neutron star surface one expects $\mu\equiv 1$, but the force-free region is separated from the stellar surface by the sheet of current closure and pair formation. As is shown in \cite{p16}, the step in $\mu$ across the sheet is indeed consistent with the moment of the closing current dictated by the magnetosphere.

\begin{figure*}
\includegraphics[width=150mm]{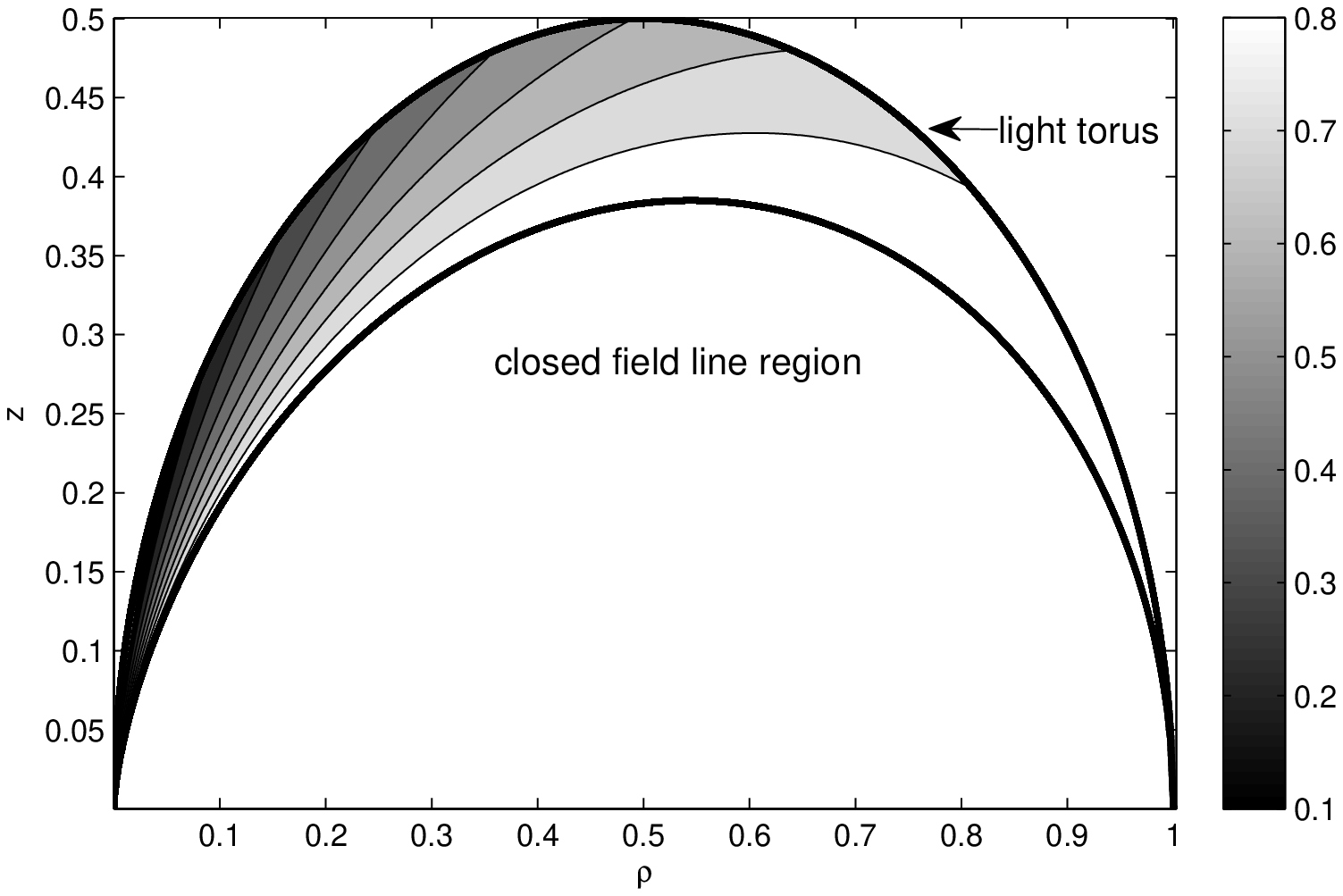}
\caption{Geometry of the force-free region considered in the paper (see text for details). Bold lines show the light torus surface and the boundary of the open field line region. Also plotted are the magnetic field lines corresponding to the magnetic flux function levels from 0.1 to 0.9 with the step 0.1.}
\label{f1}
\end{figure*}

As $\mu$ changes with $f$, the characteristic surface, where the linear velocity of rotation equals the speed of light, $\mu r\sin\theta=1$, is no longer the light cylinder. With $\mu(f)$ given by Eq.~(\ref{eq2}), the light surface is described by the equation $r=C_1^{-1/3}\sin\theta$ and represents the torus whose small and large radii are equal (see Fig.~\ref{f1} and 
\cite{p16} for details).

Equilibrium condition for the separatrix \cite{o74,l90,u03} and its stability \cite{c05} dictate the choice of constants $C_1=1$ and $C_2=-4$ \cite{p16}. Note that in the closed field line region, where $g\equiv 0$ and $\mu\equiv 1$, the force-free flux function should differ from the purely dipolar one. In the open field line region, for the solution (\ref{eq1})-(\ref{eq2}) we have $B^2-E^2>0$ everywhere, except for the Y-point $(r=1,\theta=\pi/2)$, where it turns into zero. Therefore we assume termination of the force-free regime and current closure at the light torus, in which case in the axial region, where both $g$ and $\mu$ given by Eq.~(\ref{eq2}) show discontinuity, the force-free regime is excluded naturally.

With such a general magnetospheric picture in hand, in the present paper we consider the segment of the force-free region lying between the light torus surface and the closed field line zone (see Fig.~\ref{f1}). Most probably, it is the place of generation of the pulsar radio and non-thermal high-energy emissions. Let us also note in passing that the light torus surface itself is suggestive as a very-high-energy emission region. On the whole, all the boundaries of the region considered (i.e. the pair formation front close to the neutron star, the separatrix bounding the closed field line zone and the light torus surface with the closing current sheet) are rich in physics. Thus, our present study is believed to provide a basis for diversiform advances in pulsar research.

\section{Two-fluid model in the force-free approximation}
First let us address the self-consistent two-fluid model of the pulsar magnetosphere in the force-free approximation. In application to the monopolar configuration of the magnetic field, this was already done in \cite{p15}. Here we generalize that treatment to the case of arbitrary force-free fields in order to derive some general features and make our formalism applicable to the purely dipolar case.

In the spherical coordinate system $(r,\theta,\phi)$ with the axis along the pulsar axis, the axisymmetric force-free magnetic and electric field strengths, $\mathbf{B}$ and $\mathbf{E}$, can be presented as
\begin{equation}
\mathbf{B}=\frac{\nabla f\times\mathbf{e}_\phi}{r\sin\theta}+\frac{g}{r\sin\theta}\mathbf{e}_\phi\,,\quad\mathbf{E}=-\mu\nabla f\,.
\label{eq3}
\end{equation}
With the electromagnetic fields in the form (\ref{eq3}), the Maxwell's equations $\nabla\cdot \mathbf{B}=0$ and $\nabla\times\mathbf{E}=0$ as well as the ideality condition $\mathbf{E}\cdot\mathbf{B}=0$ are fulfilled automatically. The two-fluid description also involves the other two Maxwell's equations,
\begin{equation}
n_+\mathbf{v}_+-n_-\mathbf{v}_-=\nabla\times\mathbf{B}\,,
\label{eq4}
\end{equation}
\begin{equation}
n_+-n_-=\nabla\cdot\mathbf{E},
\label{eq5}
\end{equation}
(where $n_\pm$ are the electron and positron number densities and $\mathbf{v}_\pm$ the corresponding velocities), the continuity equations for the two particle species,
\begin{equation}
\nabla\cdot\left (n_\pm\mathbf{v}_\pm\right )=0\,,
\label{eq6}
\end{equation}
and the force-free equation of motion,
\begin{equation}
\mathbf{E}+\mathbf{v}_\pm\times\mathbf{B}=0\,.
\label{eq7}
\end{equation}
In the latter equation, the effects of particle inertia, pressure and gravitation are neglected. Furthermore, although the radiation damping is generally strong, it does not contribute to the force-free approximation (see below) and, correspondingly, the radiation reaction force is also ignored.

Because of the axisymmetry of the problem, we have $(\nabla f,\mathbf{e}_\phi)=0$, and the scalar product of the equation of motion (\ref{eq7}) by $\mathbf{e}_\phi$ yields
\begin{equation}
\left (\mathbf{v}_\pm,\nabla f\right )=0.
\label{eq8}
\end{equation} 
Thus, the poloidal component of the particle velocity, $\mathbf{v}_{p\pm}$, is directed along the magnetic field lines, precluding radiation damping. The explicit expression for the poloidal velocity can be obtained by taking the scalar product of the equation of motion (\ref{eq7}) by $\nabla f$ and keeping in mind Eq.~(\ref{eq8}),
\begin{equation}
\mathbf{v}_{p\pm}=\nabla f\times \mathbf{e}_\phi\frac{v_{\phi\pm}-\mu r\sin\theta}{g}\,,
\label{eq9}
\end{equation}
where $v_{\phi\pm}$ are the azimuthal components of the particle velocities.

With the magnetic field in the form (\ref{eq3}), the poloidal current density reads
\begin{equation}
\left (\nabla\times\mathbf{B}\right )_p=\frac{\nabla g\times\mathbf{e}_\phi}{r\sin\theta}\,.
\label{eq10}
\end{equation} 
Substituting Eqs.~(\ref{eq9}) and (\ref{eq10}) into the poloidal component of Eq.~(\ref{eq4}) and taking into account that for the multipolar magnetic fields $(\nabla\times\mathbf{B})_\phi\equiv 0$ yields
\begin{equation}
n_+-n_-=-\frac{1}{\mu r^2\sin^2\theta}g\frac{\mathrm{d}g}{\mathrm{d}f}\,.
\label{eq11}
\end{equation}
On the other hand, from Eq.~(\ref{eq5}) we have
\begin{equation}
n_+-n_-=-\mu\Delta f-\frac{\mathrm{d}\mu}{\mathrm{d}f}\left (\nabla f\right )^2\,.
\label{eq12}
\end{equation}
Combining Eqs.~(\ref{eq11}) and (\ref{eq12}), we arrive at the pulsar equation for the case of multipolar magnetic fields.

From the azimuthal component of Eq.~(\ref{eq4}) and Eq.~(\ref{eq5}) one can obtain the particle number densities,
\begin{equation}
n_\pm=\frac{v_{\phi\mp}\nabla\cdot\mathbf{E}}{v_{\phi-}-v_{\phi+}}\,.
\label{eq13}
\end{equation}
And finally, making use of Eqs.~(\ref{eq5}), (\ref{eq9}), (\ref{eq11}) and (\ref{eq13}) in the continuity equation (\ref{eq6}), one can find that
\begin{equation}
\frac{w_+}{w_-}=\eta(f)\,,
\label{eq14}
\end{equation}
where $w_\pm\equiv(v_{\phi\pm}-\mu r\sin\theta)/v_{\phi\pm}$ and $\eta$ is an arbitrary function of $f$.

Note that if one of the functions $v_{\phi\pm}$ were known, the other one would be given by Eq.~(\ref{eq14}), in which case the poloidal velocities (\ref{eq9}) and number densities (\ref{eq13}) would be determined as well. As the force-free approximation actually leaves an uncertainty in the particle characteristics, it is the low-mass limit of the two-fluid model which is believed to further constrain the force-free quantities.

\section{Low-mass limit of the two-fluid model}
Taking into account the particle inertia, the equation of motion is written as
\begin{equation}
\xi\left (\mathbf{v}_\pm\cdot\nabla\right )\gamma_\pm\mathbf{v}_\pm=\pm\left (\mathbf{E}+\mathbf{v}_\pm\times\mathbf{B}\right )\,,
\label{eq15}
\end{equation}
where $\gamma_\pm\equiv(1-v_\pm^2)^{-1/2}$ are the Lorentz-factors of the two particle species and $\xi\propto m$ is the numerical factor,
\begin{equation}
\xi\equiv\frac{2\Omega_*}{\omega_{G_L}}\,,
\label{eq16}
\end{equation}
with $\omega_{G_L}\equiv eB_L/(mc)$ being the particle gyrofrequency and $B_L$ the magnetic field strength at a distance $c/\Omega_*$ from the neutron star (for details see \cite{p15}).

The numerical estimate of $\xi$ reads
\begin{equation}
\xi=10^{-7}P^2\left (\frac{B_*}{10^{12}\,\mathrm{G}}\right )^{-1}\left (\frac{R_*}{10^6\,\mathrm{cm}}\right )^{-3}\,,
\label{eq17}
\end{equation}
where $P$ is the pulsar period in seconds, $B_*$ the magnetic field strength at the neutron star surface and $R_*$ the stellar radius. One can see that the left-hand side of the equation of motion (\ref{eq15}) is small on condition
\begin{equation}
\xi\gamma_c\ll 1\,,
\label{eq18}
\end{equation}
where $\gamma_c$ is the characteristic Lorentz-factor of the particles. Although the current models of pulsar gamma-ray emission incorporate high enough Lorentz-factors, $\gamma_c\sim 10^5-10^6$ \cite{ck10,g13,hk15}, the inequality (\ref{eq18}) is satisfied for typical pulsar parameters. Hence, the quantities entering the equation of motion (\ref{eq15}) can be presented as
\[
\mathbf{v}_\pm=\mathbf{v}_{0\pm}+\xi\mathbf{v}_{1\pm}+\dots\,,
\]
\[
\mathbf{E}_\pm=\mathbf{E}_{0\pm}+\xi\mathbf{E}_{1\pm}+\dots\,,
\]
\begin{equation}
\mathbf{B}_\pm=\mathbf{B}_{0\pm}+\xi\mathbf{B}_{1\pm}+\dots\,,
\label{eq19}
\end{equation}
where $\mathbf{v}_{0\pm}$, $\mathbf{E}_{0\pm}$ and $\mathbf{B}_{0\pm}$ are the force-free quantities involved into the formalism of the preceding section; hereafter the subscripts '0' are omitted. Below we consider the first approximation in $\xi$ in order to further restrict the zeroth-order force-free quantities from the solvability condition for the first-order ones.

Strictly speaking, the force balance in the equation of motion should be supplemented with the radiation reaction force \cite{ll71}
\[
\mathbf{F}_\mathrm{rad}=\xi\alpha\gamma\left [\left (\mathbf{v}\cdot\nabla\right )\mathbf{E}+\mathbf{v}\times\left (\mathbf{v}\cdot\nabla\right )\mathbf{B}\right ] \]
\[
+\alpha\left [\mathbf{E}\times\mathbf{B}+\mathbf{B}\times\left (\mathbf{B}\times\mathbf{v}\right )+\mathbf{E}\left (\mathbf{v}\cdot\mathbf{E}\right )\right ]
\]
\begin{equation}
-\alpha\gamma^2\mathbf{v}\left [\left (\mathbf{E}+\mathbf{v}\times\mathbf{B}\right )^2-\left (\mathbf{v}\cdot\mathbf{E}\right )^2\right ]\,,
\label{eq20}
\end{equation}
where the relevant normalization of the quantities is already introduced, $\alpha\equiv r_e\omega_{G_L}/c$, $r_e\equiv e^2/(mc^2)$ is the classical electron radius and, correspondingly, $\alpha\ll 1$. One can see that for the force-free quantities obeying Eq.~(\ref{eq7}) all the three terms in Eq.~(\ref{eq20}) are identically zero. This signifies that the equation $\mathbf{F}_\mathrm{rad}=0$ has a more general solution than that generally known, where the particles merely slide along the straight field lines at a speed of light. Thus, the force-free approximation generally takes into account $\mathbf{F}_\mathrm{rad}$ self-consistently.

To the first order in $\xi$, the first and third terms of $\mathbf{F}_\mathrm{rad}$ are absent as well. Hence, hereafter we consider only the second term of the radiation reaction force. At the conditions relevant to the pulsar magnetosphere, the radiation damping can be strong, especially close to the neutron star surface (e.g., \cite{f89,gruz08}). In our case, it can be quantified as
\begin{equation}
\frac{\alpha_*}{\xi\gamma_c}=10^3P^{-2}\gamma_c^{-1}\left (\frac{B_*}{10^{12}\,\mathrm G}\right )^2\,,
\label{eq21}
\end{equation}
where $\alpha_*$ corresponds to the stellar surface. As is shown in Sect.~4.4 below, it is impossible to construct the self-consistent two-fluid model of a force-free dipole without taking into account the radiation damping.

\subsection{Force balance along the magnetic field}
Let us turn to the equation of motion linearized in $\xi$ and consider its projection onto the force-free magnetic field. Keeping in mind the above considerations and making use of the force-free equation of motion (\ref{eq7}), the linearized radiation reaction force can be written as

\[\mathbf{F}_{\mathrm{rad}_1}
=\alpha\left (\mathbf{E}_1+\mathbf{v}\times\mathbf{B}_1+\mathbf{v}_1\times\mathbf{B}\right )\times\mathbf{B}\]
\begin{equation}
+\alpha\mathbf{E}\left (\mathbf{v}_1\cdot\mathbf{E}+\mathbf{v}\cdot\mathbf{E}_1\right )\,.
\label{eq22}
\end{equation}
Then, taking into account the force-free ideality condition $\mathbf{E}\cdot\mathbf{B}\equiv 0$, one can see that $\mathbf{F}_{\mathrm{rad}_1}\cdot\mathbf{B}\equiv 0$.
The scalar product of the linearized Lorentz force by $\mathbf{B}$ reads $\mathbf{F}_{\mathrm{L}_1}=\mathbf{E}_1\cdot\mathbf{B}+\mathbf{E}\cdot\mathbf{B}_1$, where the right-hand side can be recognized as the first-order longitudinal accelerating electric field. The inertial term of the equation of motion can also be simplified proceeding from the equality
\[
\left (\mathbf{v}\cdot\nabla\right )\gamma\mathbf{v}=-\mathbf{v}\times\left (\nabla\times\gamma\mathbf{v}\right )+\nabla\gamma
\]
and applying the vector identity
\[
\nabla\cdot\left (\mathbf{x}\times\mathbf{y}\right )=\mathbf{y}\times\left (\nabla\times\mathbf{x}\right )-\mathbf{x}\times\left (\nabla\times\mathbf{y}\right )\,,
\]
where $\mathbf{x}$ and $\mathbf{y}$ are arbitrary vectors. After some manipulation, the longitudinal projection of the linearized equation of motion finally takes the form
\begin{equation}
\mathbf{B}_\mathrm{p}\cdot\nabla\left [\gamma_\pm\left (1-\mu r\sin\theta v_{\phi_\pm}\right )\right ]=\pm\left (\mathbf{E}\cdot\mathbf{B}_1+\mathbf{B}\cdot\mathbf{E}_1\right )\,.
\label{eq23}
\end{equation}

If the first-order longitudinal electric field were zero, the expression in the square brackets in Eq.~(\ref{eq23}) would be a function of $f$ and could be written as
\begin{equation}
\frac{w_\pm+a}{\sqrt{(1-b)w_\pm^2+2w_\pm+a}}=C_\pm(f)\,,
\label{eq24}
\end{equation}
where $a\equiv 1-\mu^2r^2\sin^2\theta$, $b\equiv\mu^2r^2\sin^2\theta(\nabla f)^2/g^2$ and $C_\pm(f)$ signify the initial Lorentz-factor distributions at $r\to 0$ for the two particle species. The solution of Eq.~(\ref{eq24}) with respect to $w_\pm$ reads
\begin{equation}
w_\pm\equiv w_0=\frac{C_\pm^2-a\pm C_\pm\sqrt{\left (C_\pm^2-a\right )(1-a+ab)}}{1+C_\pm^2(b-1)}\,,
\label{eq25}
\end{equation}
and in case of a force-free dipole (see Eqs.~(\ref{eq1})-(\ref{eq2}))
\begin{equation}
a=1-\frac{r^6}{\sin^6\theta}\,,\quad b=\frac{4-3\sin^2\theta}{4(1-\sin^4\theta/r^2)}\,.
\label{eq26}
\end{equation}
As $w_\pm$ are positively defined quantities, one should choose the sign '+' before the square root in Eq.~(\ref{eq25}).

As can be seen from Eq.~(\ref{eq5}) and the azimuthal component of Eq.~(\ref{eq4}), the azimuthal velocity of the two particle species should be different. At the same time, the velocities obtained from Eq.~(\ref{eq25}) at $C_+\neq C_-$ do not obey the continuity condition (\ref{eq6}). Hence, the self-consistent two-fluid model should necessarily involve the non-zero first-order accelerating electric field.

In general, noticing that the sum of the right-hand sides of Eq.~(\ref{eq23}) for the two particle species is zero, we have
\begin{equation}
\frac{w_++a}{\sqrt{(1-b)w_+^2+2w_++a}}+\frac{w_-+a}{\sqrt{(1-b)w_-^2+2w_-+a}}
=2C(f)\,.
\label{eq27}
\end{equation}
The solution of such an equation, however, would be too complicated. Below we dwell on the physically meaningful case of a slight distinction of the electron and positron distribution functions. With the choice
\begin{equation}
w_\pm=w_0\left [1+\varepsilon\left (f\right )\right ]\,,
\label{eq28}
\end{equation}
(where $\varepsilon\ll 1$ over the whole range of $f$ considered) both the continuity condition (\ref{eq6}) and Eq.~(\ref{eq27}) are satisfied, since in the latter the terms $\sim\varepsilon$ are cancelled, whereas $w_0$ is the solution of Eq.~(24). The particle number densities then read
\begin{equation}
n_\pm=\frac{\nabla\cdot\mathbf{E}}{2}\left (\frac{1+w_0}{2\varepsilon w_0}\pm 1\right )\,,
\label{eq29}
\end{equation}
and the quantity $\varepsilon(f)$ can be interpreted as the inverse multiplicity of the plasma. Thus, for given initial Lorentz-factor $\gamma_i\equiv C(f)$ and multiplicity $\varepsilon^{-1}(f)$ of the plasma particles the force-free velocity and number density distributions of the two particle species are completely determined. As in our consideration $\xi\ll\varepsilon\ll 1$ and $\xi$ is typically negligible (see Eq.~(\ref{eq17})), the force-free distributions are believed to be a proper representation of the plasma flow characteristics in the region considered.

\begin{figure*}
\vspace{-2cm}
\includegraphics[width=110mm]{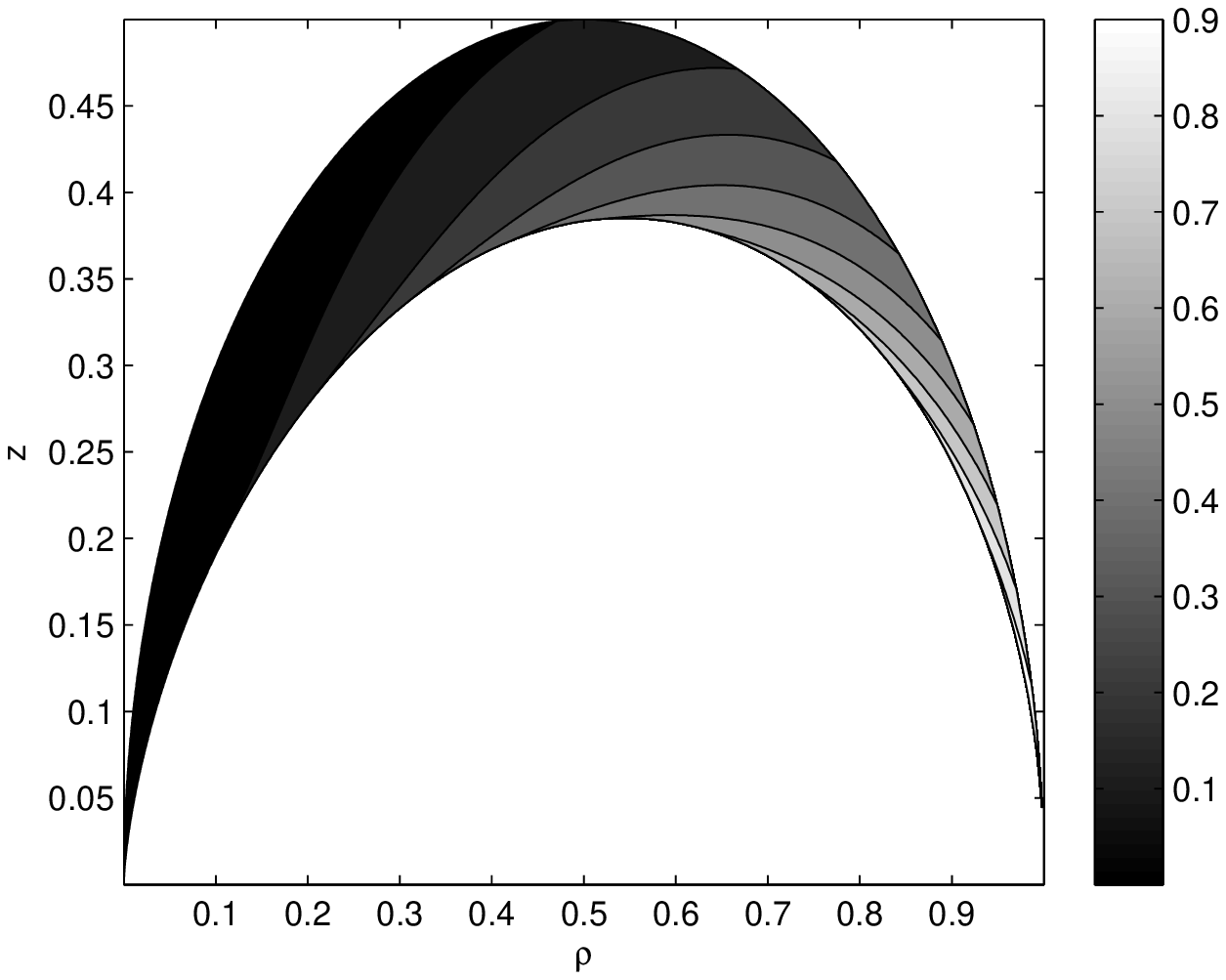}
\includegraphics[width=110mm]{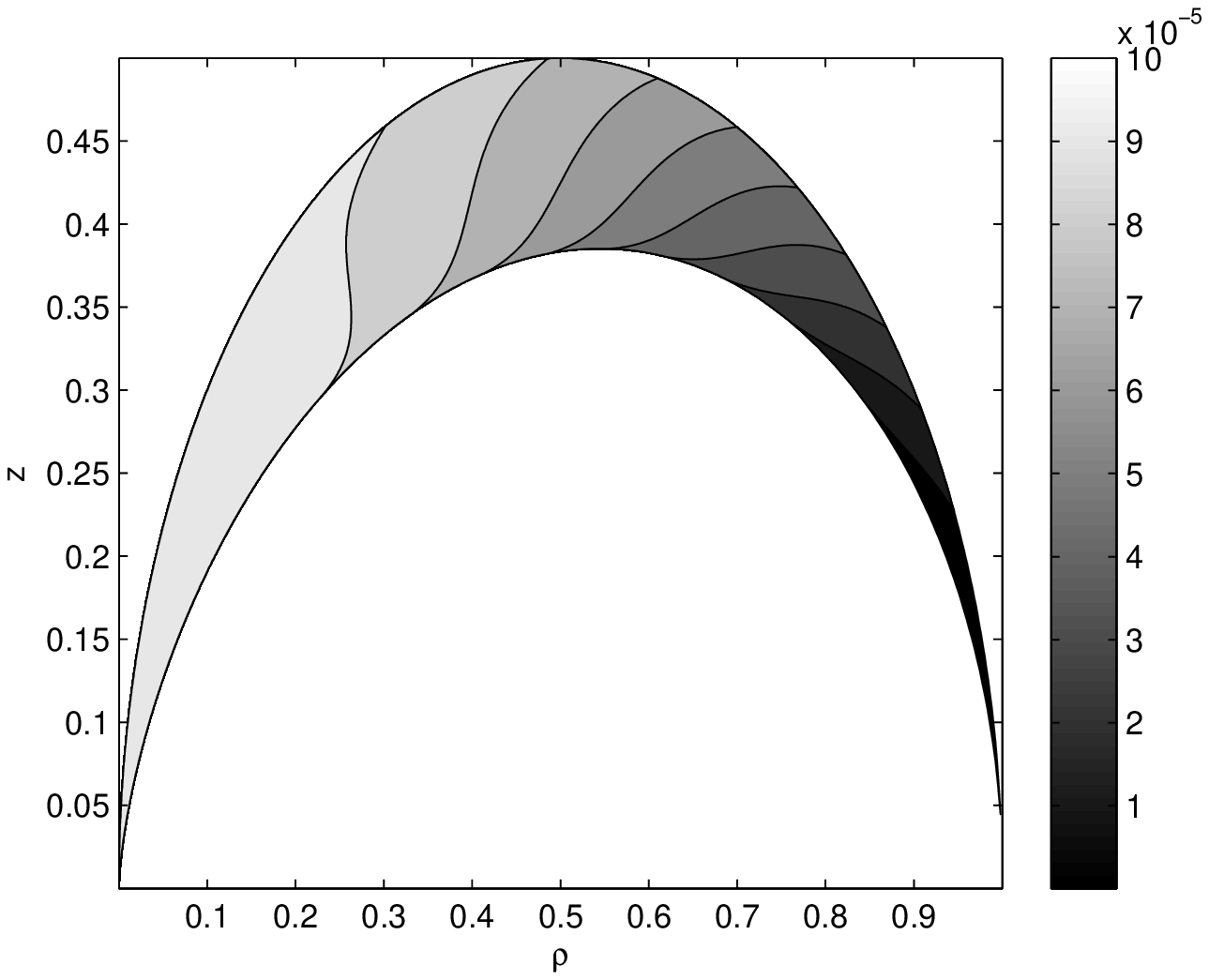}
\includegraphics[width=110mm]{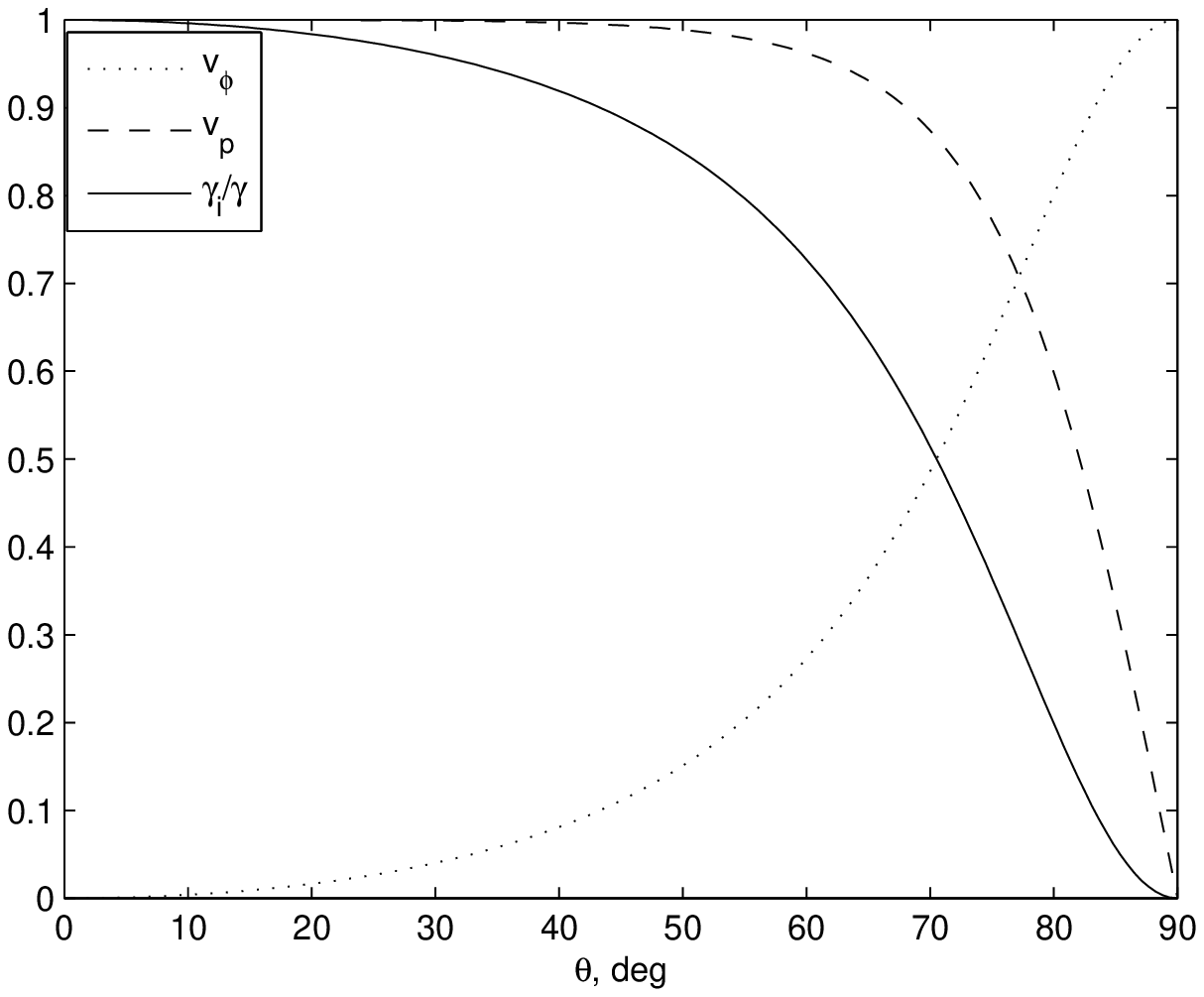}
\caption{Velocity characteristics of the particles sustaining dipolar force-free configuration of the pulsar magnetosphere, $\gamma_i=100$; a) distribution of the azimuthal velocity $v_\phi$; b) distribution of the inverse square of the Lorentz-factor, $1-v_p^2-v_\phi^2$; c) poloidal and azimuthal velocity components together with the Lorentz-factor at the light torus as functions of the polar angle.}
\label{f2}
\end{figure*}

Figure \ref{f2},a shows the isolines of the quantity $v_\phi\equiv \mu r\sin\theta /(1+w_0)$ in the force-free region of the dipolar magnetosphere of a pulsar. The corresponding distribution of the inverse square of the Lorentz-factor, $1-v_\phi^2-v_p^2$, is plotted in Fig.~\ref{f2},b. The quantities $v_\phi$, $v_p$ and $\gamma_i/\gamma$ at the light torus surface ($r=\sin\theta$) are presented in Fig.~\ref{f2},c as functions of the polar angle $\theta$. Note an increase of $v_\phi$ and $\gamma$ with distance from the neutron star and from the magnetic axis. Over most part of the region considered, $v_\phi$ remains small and $\gamma/\gamma_i\leq 10$, whereas in the vicinity of the Y-point both quantities increase drastically (cf. Fig.~\ref{f2},c) with $v_\phi\to 1$ and $\gamma\to\infty$.

\begin{figure*}
\includegraphics[width=150mm]{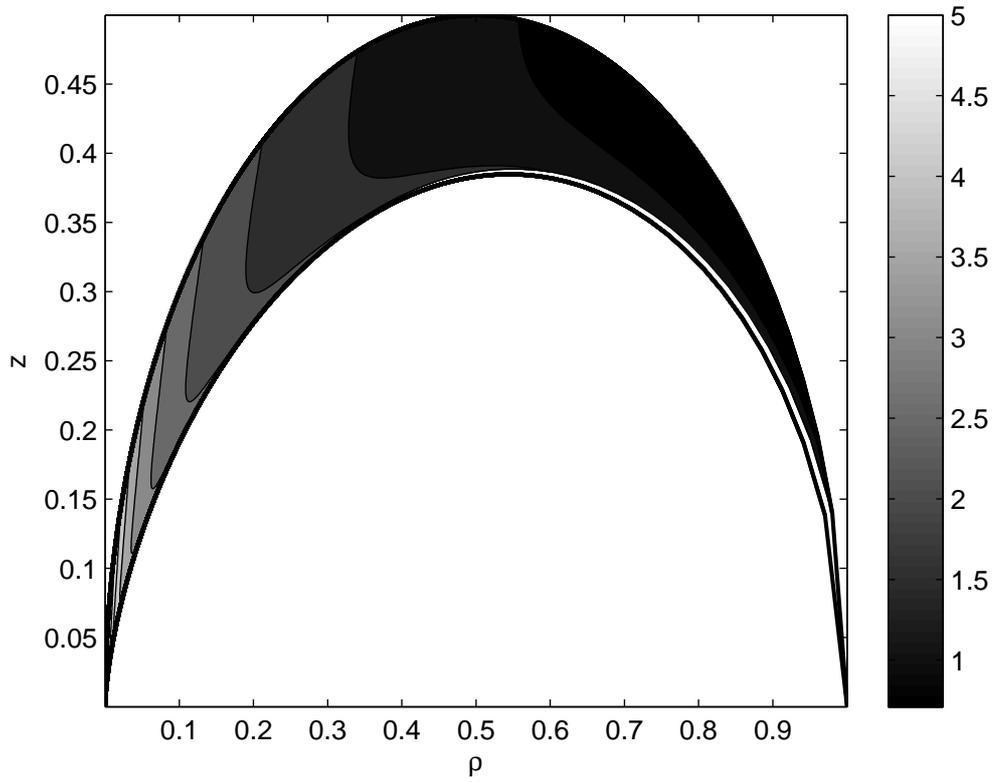}
\includegraphics[width=150mm]{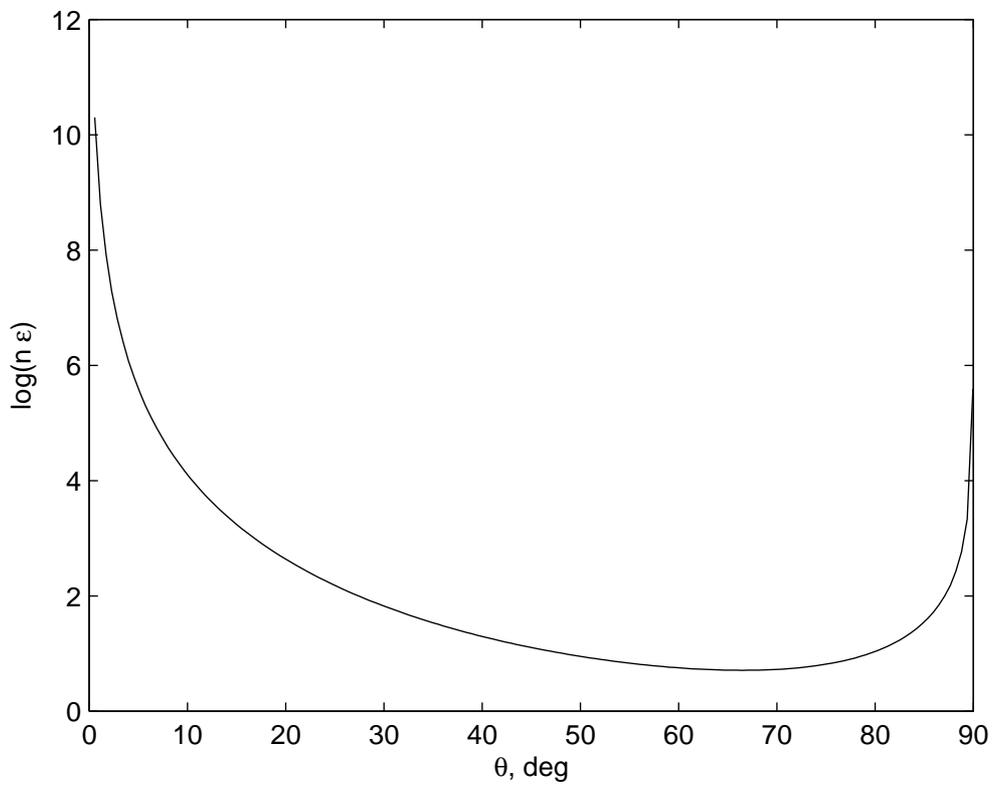}
\caption{Number density of the particles sustaining dipolar force-free configuration of the pulsar magnetosphere, $\gamma_i=100$; a) distribution of $\log(n\varepsilon)$; b) $\log(n\varepsilon)$ at the light torus surface as a function of the polar angle.}
\label{f3}
\end{figure*}

The isolines of the logarithm of the particle number density are shown in Fig.~\ref{f3},a and its distribution at the light torus in Fig.~\ref{f3},b. One can see that the number density is large not only close to the neutron star but also in the vicinity of the Y-point. Thus, the latter region is suggestive as a site of the pulsar very-high-energy emission. Note that for display purposes, in Figs.~\ref{f2},\ref{f3} we have presented the 'effective' quantities based on $w_0$, keeping in mind that the actual distributions of the two particle species differ slightly.

\subsection{Particle trajectories}
With the force-free velocities in hand, one can find the particle trajectories from the standard set of equations
\[
\frac{\mathrm{d}r_\pm}{\mathrm{d}t}=v_{r_\pm}\,,
\]
\[
r_\pm\frac{\mathrm{d}\theta_\pm}{\mathrm{d}t}=v_{\theta_\pm}\,,
\]
\begin{equation}
r_\pm\sin\theta_\pm\frac{\mathrm{d}\phi_\pm}{\mathrm{d}t}=v_{\phi_\pm}\,,
\label{eq30}
\end{equation}
where $t$ can be regarded as a parameter along the trajectory. Excluding $t$ from the first and second equations of the set (\ref{eq30}) and making use of Eq.~(\ref{eq9}) yield
\begin{equation}
\frac{\mathrm{d}r_\pm}{\mathrm{d}\theta_\pm}=-\frac{\partial f/\partial\theta}{\partial f/\partial r}\,.
\label{eq31}
\end{equation}
Thus, in the poloidal plane the electron and positron trajectories exactly follow the magnetic field lines. In the dipolar case, Eq.~(\ref{eq31}) results in $\sin^2\theta_\pm=fr_\pm$.

\begin{figure*}
\includegraphics[width=150mm]{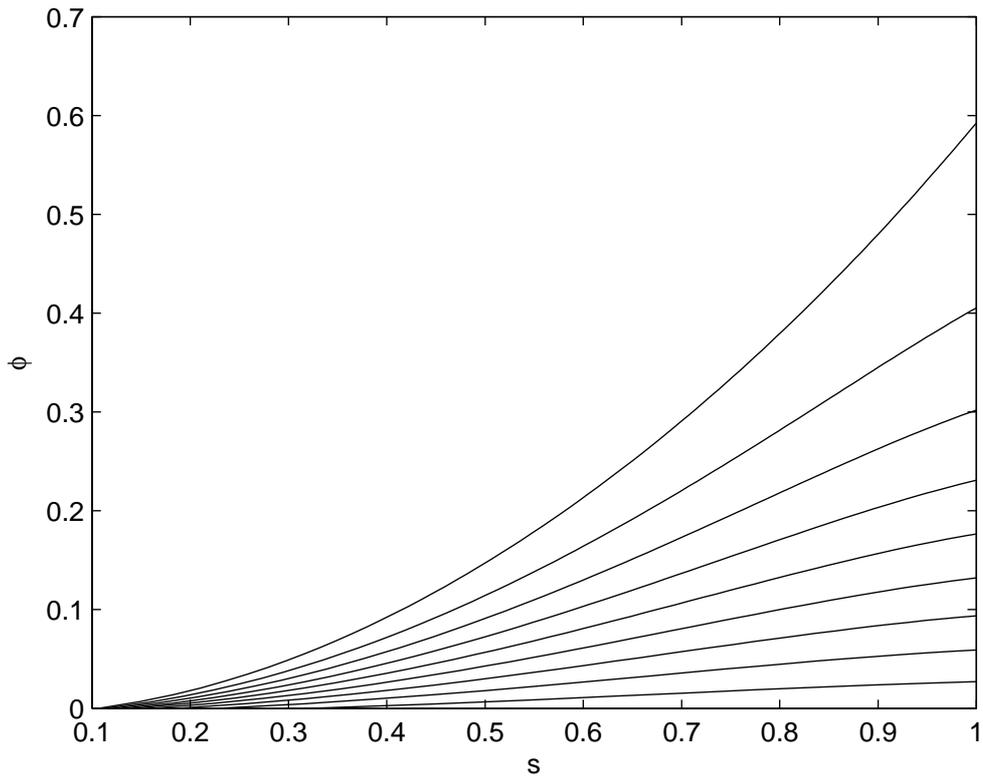}
\caption{Particle azimuthal coordinate as a function of the parameter $s$ for different magnetic field lines, $\gamma_i=100$ (see text for details); the curves (from bottom to top) correspond to the flux function values from 0.1 to 0.9 with the step 0.1.}
\label{f4}
\end{figure*}

From the first and the third equations of the set (\ref{eq30}) one finds
\begin{equation}
\mathrm{d}\phi_\pm =\frac{\mathrm{d}r_\pm}{2f\sqrt{1-fr_\pm}}\frac{g(f)}{w_\pm (f,r)}\,,
\label{eq32}
\end{equation}
where $w_\pm$ is given by Eqs.~(\ref{eq25}), (\ref{eq28}) and is expressed in terms of $f,r$. The results of numerical integration of Eq.~(\ref{eq30}) with $w_0$ instead of $w_\pm$ for different $f$ in the dipolar case are presented in Fig.~{\ref{f4}}, where the azimuthal coordinate is shown as a function of $s\equiv r/\sin\theta$. The quantity $s$ is the parameter along a magnetic field line and is defined as $s\equiv S/S_L$, where $S\equiv \sqrt{\cos\theta}/r$ satisfies the condition $\nabla S\cdot\nabla f\equiv 0$ and takes the value $S_L=\sqrt{\cos\theta}/\sin\theta$ at the light torus surface $r=\sin\theta$. As can be seen form Fig.~\ref{f4}, stronger variation of $\phi$ corresponds to larger $f$. However, for any $f$ the azimuthal twist of the particle trajectory is much weaker than that of the force-free magnetic line. This is analogous to the case of a force-free monopole (cf. figure 5 in \cite{p15}). Recall that according to the classical result of \cite{m73} the 'massless' particles strictly follow the poloidal magnetic lines without any azimuthal motion at all.

\subsection{Accelerating electric field}
\begin{figure*}
\includegraphics[width=150mm]{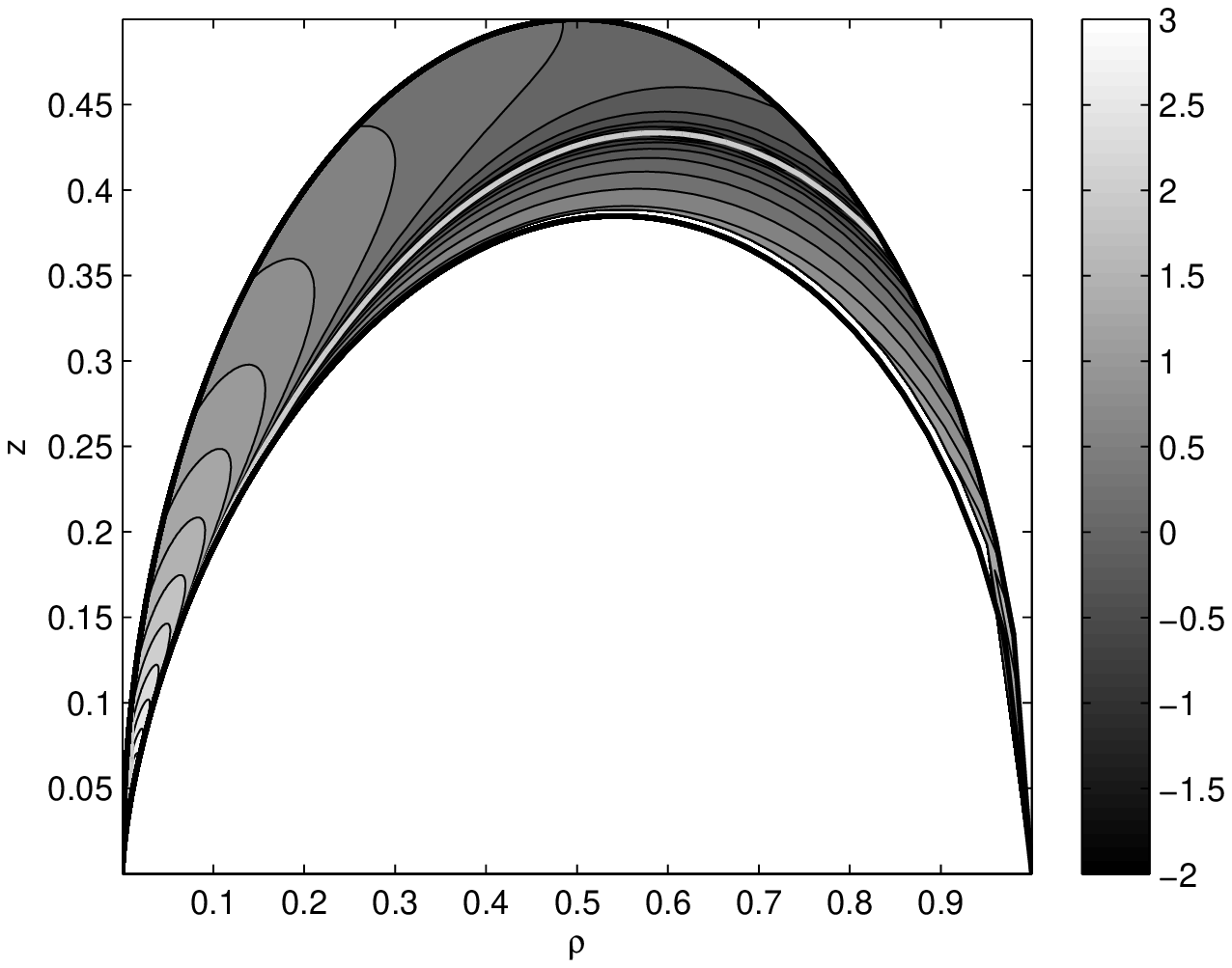}
\includegraphics[width=150mm]{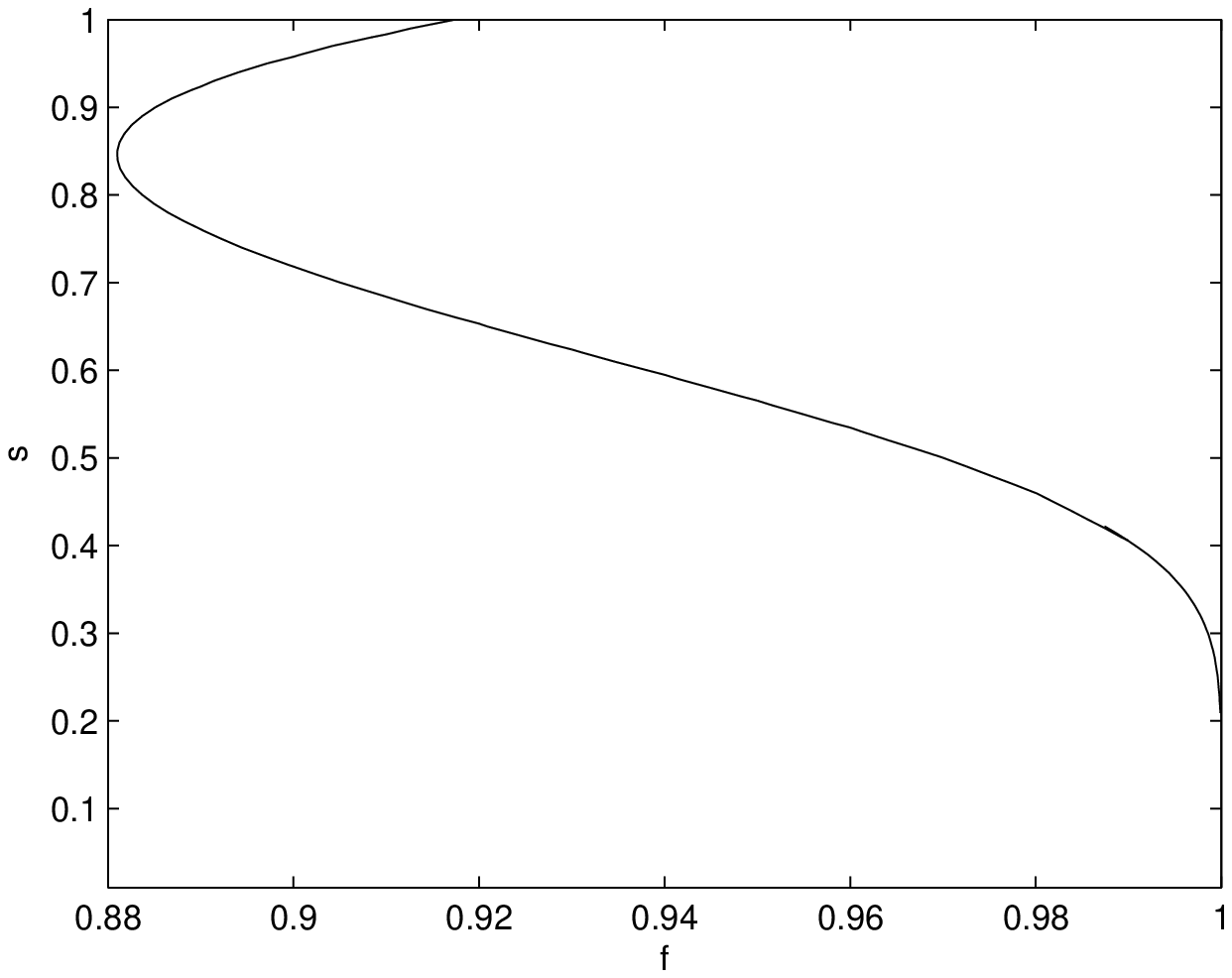}
\caption{Structure of the first-order longitudinal electric field in units of $\varepsilon\gamma_i^3$; a) distribution of $\log\vert\hat{\mathbf{E}}\cdot\hat{\mathbf{B}}\vert$ ; the line $\hat{\mathbf{E}}\cdot\hat{\mathbf{B}}=0$ is shown in light gray; b) the line $\hat{\mathbf{E}}\cdot\hat{\mathbf{B}}=0$ in the coordinates $(f,s)$.}
\label{f5}
\end{figure*}

Substituting Eq.~(\ref{eq28}) into Eq.~(\ref{eq23}) and retaining the terms $\sim\varepsilon$ yield the longitudinal electric field in the form
\begin{equation}
\mathbf{E}_1\cdot\mathbf{B}+\mathbf{E}\cdot\mathbf{B}_1=\varepsilon\gamma_i^3(f)\mathbf{B}_p\cdot\nabla\left [\frac{w_0(ab-a+1)}{(a+w_0)^3}\right ]\,.
\label{eq33}
\end{equation}
Further manipulation using Eq.~(\ref{eq25}) leads to
\begin{equation}
\mathbf{E}_1\cdot\mathbf{B}+\mathbf{E}\cdot\mathbf{B}_1
=\varepsilon\gamma_i^3(f)\mathbf{B}_p\cdot\nabla\left [\frac{1-a/\gamma_i^2}{a}\left (1-\frac{\sqrt{1-a/\gamma_i^2}}{\sqrt{ab-a+1}}\right )\right ]\,.
\label{eq34}
\end{equation}
For the dipolar case, the isolines of the logarithm of the absolute value of the longitudinal electric field are shown in Fig.~\ref{f5},a. Note that at $f\approx 0.9$ the quantity $\mathbf{E}_1\cdot\mathbf{B}+\mathbf{E}\cdot\mathbf{B}_1$ changes the sign. The line of zero longitudinal electric field is shown in Fig.~\ref{f5},a in light gray and also in Fig.~\ref{f5},b in the coordinates $(f,s)$. As can be seen from the latter figure, for the field lines in the range $f=0.88-0.92$ the sign of the longitudinal electric field changes twice.

\subsection{Force balance across the force-free magnetic field}
To supplement the picture, it is necessary to examine the force balance across $\mathbf{B}$. Given that the radiation reaction force is neglected, the transverse projections of the resultant equation of motion immediately give the first-order velocity components perpendicular to the force-free magnetic field, $\mathbf{v}_1\cdot\mathbf{E}$ and $\mathbf{v}_1\cdot(\mathbf{E}\times\mathbf{B})$. Using them in the linearized analogue of Eq.~(\ref{eq4}) together with the force-free quantities $n_\pm$ and $\mathbf{v}_\pm$ found in Sect.~4.1 then leads to $B_1,E_1\sim 1/\varepsilon\to\infty$. Thus, the self-consistent two-fluid model of a force-free dipole cannot be constructed without taking into account some other constituents. The radiation damping seems a proper effect to be included into the model.

In case of efficient radiation damping, the transverse components of the linearized equation of motion take the form
\begin{equation}
\mathbf{F}_{L_1}\cdot\mathbf{E}=0\,,\quad \mathbf{F}_{L_1}\cdot(\mathbf{E}\times\mathbf{B})=0\,,
\label{eq35}
\end{equation}
(where $\mathbf{F}_{L_1}$ is the linearized Lorentz force) and after some manipulation are reduced to
\begin{equation}
\mathbf{v}_1\cdot\mathbf{E}+\mathbf{v}\cdot\mathbf{E}_1=\frac{\mathbf{v}\cdot\mathbf{B}(\mathbf{E}\cdot\mathbf{B}_1+\mathbf{E}_1\cdot\mathbf{B})}{B^2-E^2}\,,
\label{eq36}
\end{equation}
\begin{equation}
\mathbf{F}_{L_1}\cdot\mathbf{E}=0\,.
\label{eq37}
\end{equation}

Combining Eq.~(\ref{eq4}) with its linearized counterpart and incorporating Eqs.~(\ref{eq36}), (\ref{eq37}) yield, respectively,
\begin{equation}
(\mathbf{E}+\xi\mathbf{E}_1)\cdot\nabla\times((\mathbf{B}+\xi\mathbf{B}_1))=\xi(\mathbf{E}_1\cdot\mathbf{B}+\mathbf{E}\cdot\mathbf{B}_1)\frac{\mathbf{B}\cdot\nabla\times\mathbf{B}}{B^2-E^2}\,,
\label{eq38}
\end{equation}
\begin{equation}
(\mathbf{E}_1\cdot\mathbf{B}+\mathbf{E}\cdot\mathbf{B}_1)\cdot\nabla\times\mathbf{B}-E^2\nabla\cdot\mathbf{E}_1=(\mathbf{B}\times\mathbf{E})\cdot\nabla\times\mathbf{B}_1\,.
\label{eq39}
\end{equation}
In terms of the total fields $\hat{\mathbf{B}}\equiv\mathbf{B}+\xi\mathbf{B}_1$ and $\hat{\mathbf{E}}\equiv\mathbf{E}+\xi\mathbf{E}_1$, Eq.~(\ref{eq39}) is compactly rewritten as
\begin{equation}
\mathbf{E}\cdot\left \{\hat{\mathbf{E}}\nabla\cdot\hat{\mathbf{E}}+(\nabla\times\hat{\mathbf{B}}\times\hat{\mathbf{B}})\right \}=0\,.
\label{eq40}
\end{equation}
The latter equation, being direct consequence of Eq.~(\ref{eq37}) and implying force balance along $\mathbf{E}$, can be recognized as the analogue of the pulsar equation for the total fields $\hat{\mathbf{B}}$ and $\hat{\mathbf{E}}$.

\subsection{Plasma conductivity}
Now we finally address the plasma conductivity in the low-mass two-fluid model of a force-free dipole. The relation between the electric field and current is generally presented as
\begin{equation}
\mathbf{j}=\sigma_0\mathbf{E}\cdot\mathbf{b}+\sigma_\mathrm{P}\mathbf{E}_\perp-\sigma_\mathrm{H}\mathbf{E}\times\mathbf{b}\,,
\label{eq41}
\end{equation}
where $\mathbf{b}$ is the unit vector along the magnetic field, $\mathbf{E}_\perp$ the electric field component perpendicular to the magnetic field, $\sigma_0$, $\sigma_\mathrm{P}$ and $\sigma_\mathrm{H}$ are the parallel, Pedersen and Hall conductivities, respectively.

In the case considered, the parallel conductivity,
\[
\sigma_0\approx\frac{\mathbf{j}\cdot\mathbf{B}}{\hat{\mathbf{E}}\cdot\hat{\mathbf{B}}}\,,
\]
takes the form
\begin{equation}
\sigma_0=\frac{(\nabla f)^2\mathrm{d}g/\mathrm{d}f}{r^2\sin^2\theta\hat{\mathbf{E}}\cdot\hat{\mathbf{B}}}\,.
\label{eq42}
\end{equation}
One can see that $\sigma_0\sim(\varepsilon\xi)^{-1}\to\infty$.

With Eq.~(\ref{eq38}), the Pedersen conductivity,
\[
\sigma_\mathrm{P}\approx\frac{\mathbf{j}\cdot\mathbf{E}}{E^2}\,,
\]
is reduced to
\begin{equation}
\sigma_\mathrm{P}=\frac{\mathbf{B}\cdot(\nabla\times\mathbf{B})\hat{\mathbf{E}}\cdot\hat{\mathbf{B}}}{E^2(B^2-E^2)}
\label{eq43}
\end{equation}
and appears as small as $(\varepsilon\xi)$.

The Hall conductivity,
\[
\sigma_\mathrm{H}=B\frac{\mathbf{j}\cdot(\mathbf{B}\times\mathbf{E})}{(\mathbf{B}\times\mathbf{E})^2}\,,
\]
reads
\begin{equation}
\sigma_\mathrm{H}=\frac{g\mathrm{d}g/\mathrm{d}f}{\mu r\sin\theta\sqrt{(\nabla f)^2+g^2}}\,.
\label{eq44}
\end{equation}
It is of order unity and is related to the particle drift in the crossed electric and magnetic fields.

\begin{figure*}
\vspace{-2cm}
\includegraphics[width=110mm]{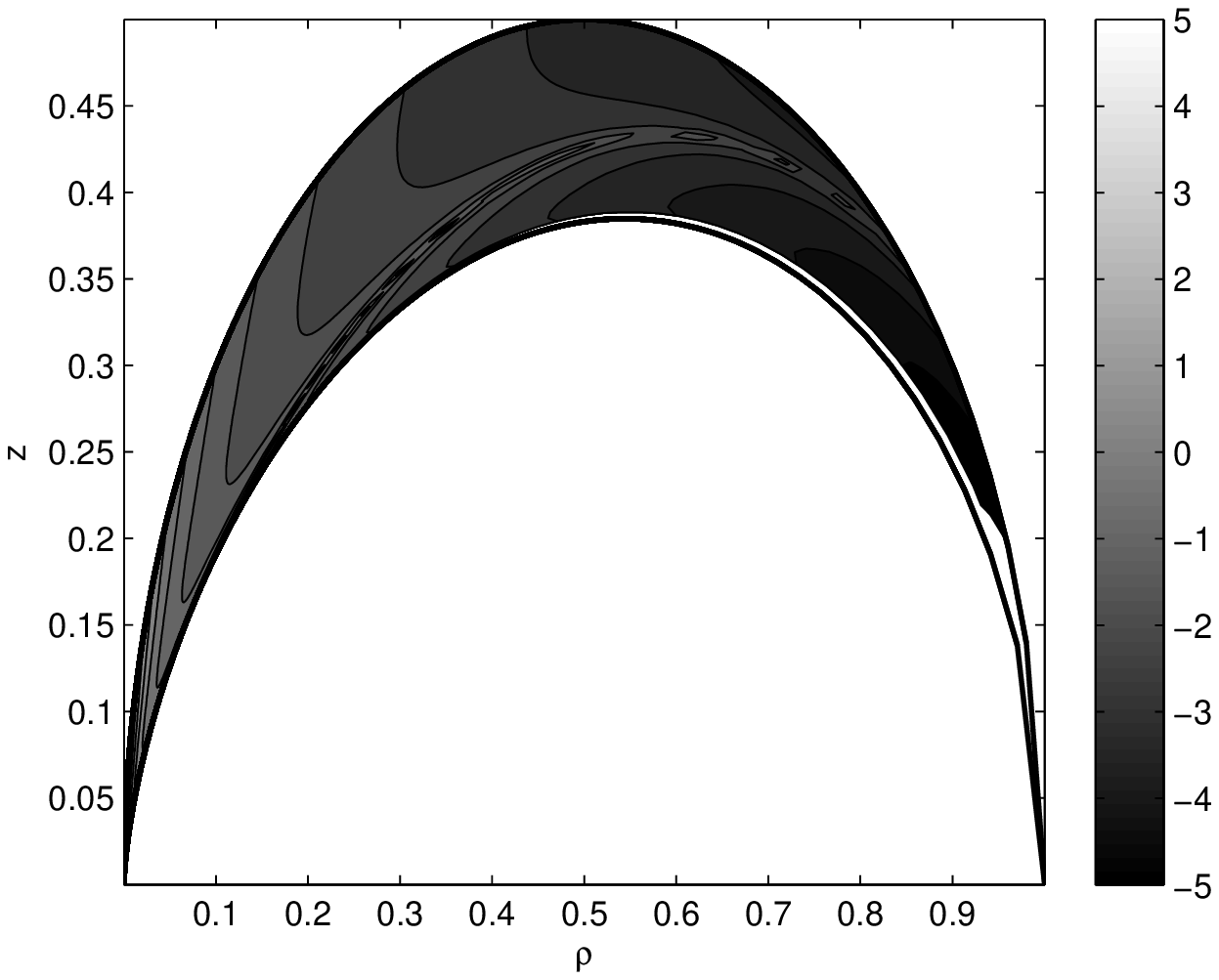}
\includegraphics[width=110mm]{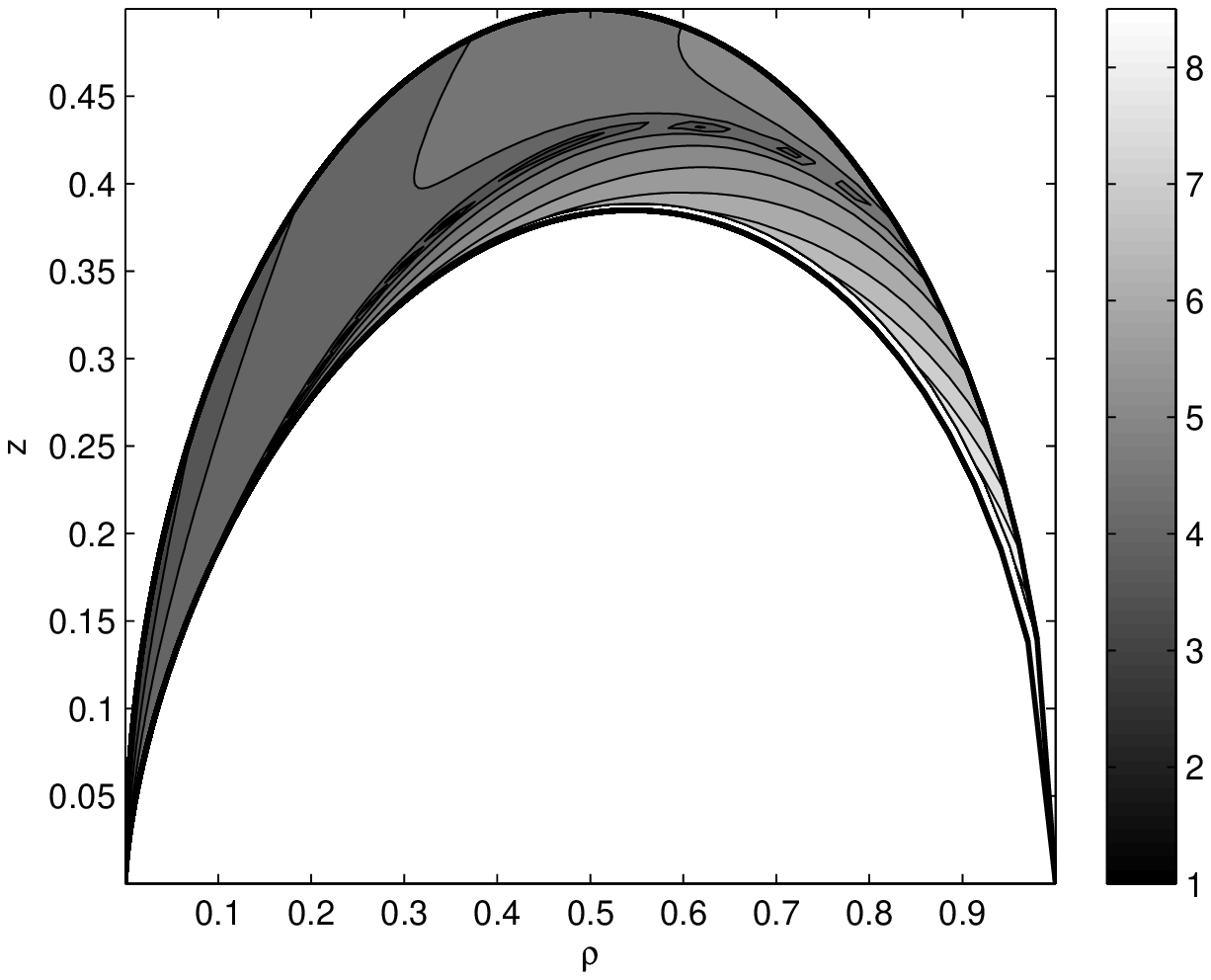}
\includegraphics[width=110mm]{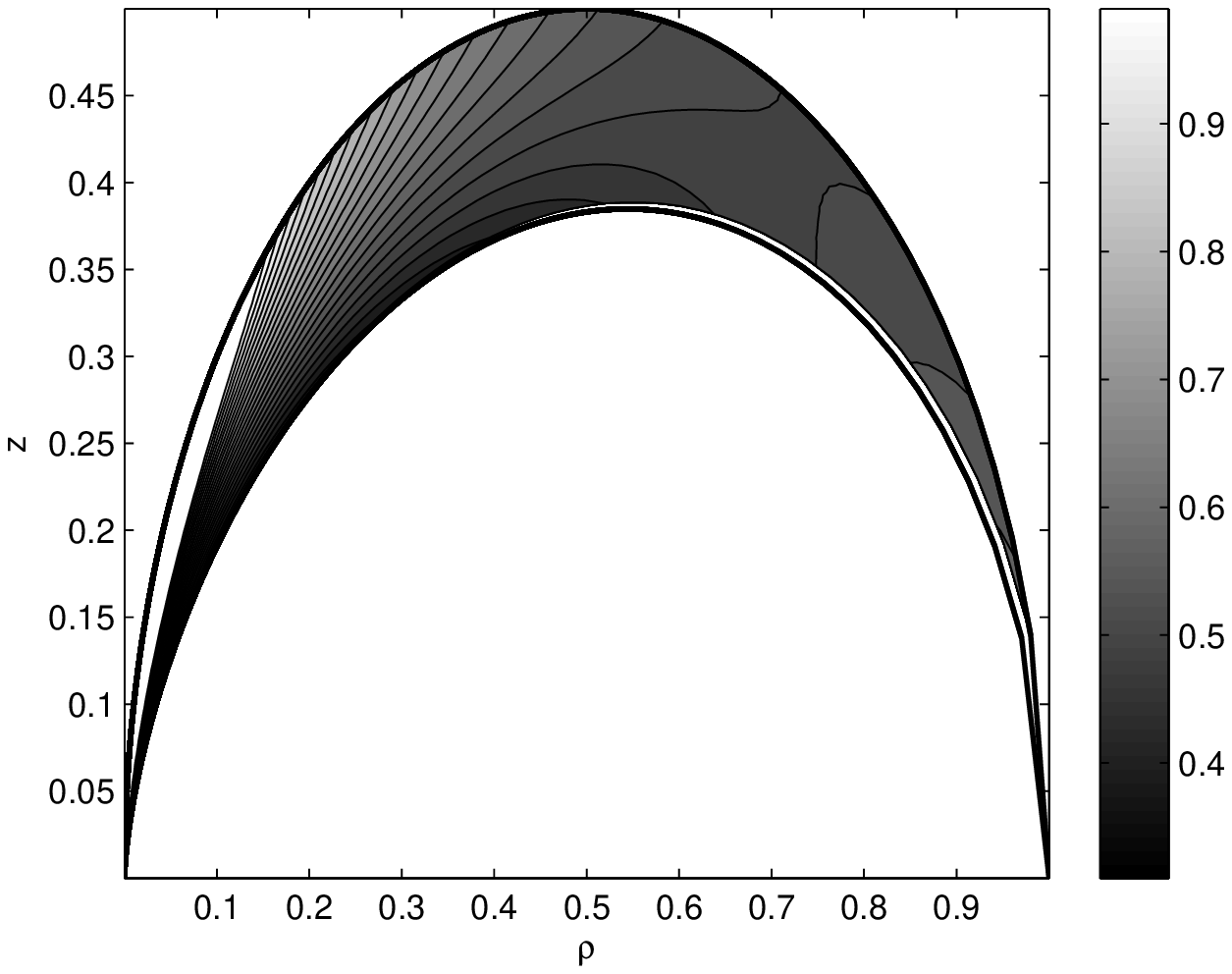}
\caption{Distributions of the plasma conductivity components, $\gamma_i=100$, $\varepsilon=0.1$; a) $\log\sigma_0$; b) $\log\sigma_P$; c) $\log\sigma_H$.}
\label{f6}
\protect\label{lastpage}
\end{figure*}

The conductivities (\ref{eq42})-(\ref{eq44}) in case of a force-free dipole are shown in Fig.~\ref{f6}. Note the complicated character of the conductivity distributions in the region considered.

\section{Discussion}
We have considered the two-fluid model of a force-free dipole with an eye to describing the physical picture in the segment of the pulsar magnetosphere enclosed between the light torus surface and the closed field line region and believed to contain copious electron-positron plasma. Starting from the force-free formalism for 'massless' particles, we have developed the first-order approximation in particle mass, $\xi\ll 1$, taking into account the radiation damping. The solvability condition for the first-order quantities allowed to further constrain the force-free distributions of the two particle species as well as to derive the particle trajectories and the components of the plasma conductivity tensor. Besides that, the first-order longitudinal electric field is found and the inertial analogue of the force-free pulsar equation is obtained.

As could be expected in advance, our present results have much in common with those found for the case of a force-free monopole (see \cite{p15}). In particular, the difference in the electron and positron velocities, $\mathbf{v}_+-\mathbf{v}_-\neq 0$ and the presence of the first-order longitudinal electric field, $\hat{\mathbf{E}}\cdot\hat{\mathbf{B}}\neq 0$, appear inherent features of the self-consistent two-fluid model. We concentrate on the physically meaningful assumption of a small velocity shear, $\vert\mathbf{v}_+-\mathbf{v}_-\vert \sim\varepsilon$ (with $\xi\ll\varepsilon\ll 1$), in which case $\hat{\mathbf{E}}\cdot\hat{\mathbf{B}}\sim\xi\varepsilon$ and $\varepsilon=\varepsilon(f)$ characterizes the inverse multiplicity of the pulsar plasma. Then for given multiplicity and Lorentz-factor distributions at the pair formation front, $\varepsilon^{-1}(f)$ and $\gamma_i(f)$, our description yields unambiguously the velocities and number densities of the two particle species at each point of the region considered.

As was already pointed out for the monopolar case, the poloidal and azimuthal velocity shear of the electrons and positrons is suggestive of the two-stream and diocotron instabilities in the pulsar plasma, which may underlie the radio emission mechanism and the subpulse drift phenomenon, respectively. Furthermore, the relation between the velocity shear and the longitudinal electric field may manifest itself as a correlation of the radio and high-energy emissions of a pulsar.

Similarly to the monopolar case, in the dipolar force-free magnetosphere the particle trajectories follow exactly the poloidal magnetic field lines, while their azimuthal twist is too small as compared to that of the field lines. At the same time, the characteristic features of the particle motion in the monopolar and dipolar cases are distinct. In the monopolar case with its cylindrical geometry, the particle motion depends only on the distance to the magnetic axis and acceleration becomes significant far from the light cylinder, at distances approximately $\gamma_i$ times larger. In the dipolar case, however, the particle motion is a function of both coordinates in the poloidal plane. Over most part of the region considered, the particle acceleration is moderate, with the Lorentz-factor $\leq 10\gamma_i$. An exception is the vicinity of the Y-point ($r=1$, $\theta=\pi/2$), where $\gamma$ grows drastically, tending to infinity, and the azimuthal velocity component becomes the dominant one. Thus, the vicinity of the Y-point seems promising as a site of the pulsar very-high-energy emission observed over the range up to 1.5 TeV \cite{vhe16}. Our result is in a general agreement with the current studies of acceleration in the pulsar magnetosphere (see, e.g., \cite{bedn12,bogoval14,hir14,hir15}).
Together with the X-type field line structure in the equatorial region beyond the light torus (for details see \cite{p16}), our picture of the particle motion close to the Y-point is suggestive of the magnetocentrifugal formation of a jet (see \cite{bp82} for general theory and \cite{bogoval14} for pulsar applications).

The longitudinal electric field structure also exhibits a substantial distinction from the monopolar case. In the dipolar magnetosphere, $\hat{\mathbf{E}}\cdot\hat{\mathbf{B}}$ appears to change sign, roughly at $f=0.9$, making the region $0.9\leq f\leq 1$ suspicious of being controlled by the outer gap. A similar behaviour of the accelerating electric field was also found in \cite{ys12} by means of particle simulation of the magnetospheric structure. Note, however, that in contrast to the previous attempts at including the outer gap into the force-free magnetosphere \cite{ys12,p13}, in our case the field lines at $0.9\leq f\leq 1$ carry direct rather than return poloidal electric current. (The same picture is also preserved for other values of $C_1/C_2$ in the current function (\ref{eq2})). Our result may well be understood if one keep in mind that the particle motion, and, in particular, acceleration, is chiefly determined by the force-free electromagnetic fields rather than by the first-order longitudinal electric field. It is the force-free fields that dictate the necessary velocity at each spatial point. Of course, our present consideration does not include the magnetospheric gaps, but it does yield the boundary conditions for such a problem and, hopefully, will facilitate its treatment in the future.

The first-oder longitudinal electric field, being proportional to the inverse multiplicity, formally determines the components of the plasma conductivity tensor. In contrast to the dissipative models assuming ad hoc zeroth-order longitudinal electric field \cite{gruz08j,kkhc12,lst12}, in our consideration the conductivity tensor is determined unambiguously. The parallel conductivity appears $\sim (\hat{\mathbf{E}}\cdot\hat{\mathbf{B}})^{-1}$, the Pedersen conductivity $\sim \hat{\mathbf{E}}\cdot\hat{\mathbf{B}}$ and the Hall conductivity $\sim (\hat{\mathbf{E}}\cdot\hat{\mathbf{B}})^0$. The former two quantities change sign together with $\hat{\mathbf{E}}\cdot\hat{\mathbf{B}}$, demonstrating that the electric current is actually determined by the force-free condition rather than by the first-order longitudinal electric field.

\section{Conclusions}
The low-mass limit of the self-consistent two-fluid model of a force-free dipole is considered. We have found the particle distributions sustaining the dipolar force-free configuration of the pulsar magnetosphere in case of high plasma multiplicity, or, equivalently, small distinction of the electron and positron distributions. The only free parameters are the initial Lorentz-factor, $\gamma_i(f)$, and multiplicity, $\varepsilon^{-1}(f)$, which are believed to be determined by the physics of the pair production cascade at the boundary of the force-free zone.

In the poloidal plane, the particles of both species move strictly along the magnetic field lines and experience moderate acceleration, $\gamma/\gamma_i\leq 10$. The azimuthal velocities in the plasma flow generally appear too low for the particles to follow the field line rotation. However, close to the Y-point the particle motion becomes predominantly azimuthal, with the Lorentz-factor growing unrestrictedly. This may underlie the pulsar very-high-energy emission and the magnetocentrifugal acceleration of a jet.

The first-order longitudinal electric field proved to be an inherent constituent of the self-consistent two-fluid model. In contrast to the monopolar case, in the region considered this quantity changes the sign. This happens roughly at $f\approx 0.9$, but the line $\hat{\mathbf{E}}\cdot\hat{\mathbf{B}}=0$ does not coincide with a field line and the lines with $f=0.88-0.92$ intersect it twice. The reverse field beyond the line $\hat{\mathbf{E}}\cdot\hat{\mathbf{B}}=0$ can be speculated to be a residual field of the outer gap. Note, however, that  the poloidal electric current has the same direction on both sides of the line $\hat{\mathbf{E}}\cdot\hat{\mathbf{B}}=0$. This can be understood if one take into account that the particle motions are chiefly determined by the force-free fields rather than by $\hat{\mathbf{E}}\cdot\hat{\mathbf{B}}$.

We have also found the components of the formal conductivity tensor relating the electric field and current in the model considered. The parallel conductivity is $\sim (\hat{\mathbf{E}}\cdot\hat{\mathbf{B}})^{-1}$, the Pedersen conductivity is $\sim \hat{\mathbf{E}}\cdot\hat{\mathbf{B}}$ and the Hall conductivity is of order unity. The low-mass analogue of the pulsar equation is derived as well.

On the whole, our present results are believed to provide a basis for extensive studies of pulsar physics both inside and outside of the region considered.

\clearpage

\clearpage

\clearpage

\clearpage

\clearpage

\clearpage

\end{document}